\documentstyle[epsfig,12pt]{article}
\begin{document}

\begin{flushright}
{\large DFUB 2000-7}
\end{flushright}

\vspace{1.5cm}

\begin{center}

{\bf Energy losses of Q-balls in Matter, Earth and Detectors.}\\
\vspace{1.5cm}
{\bf  Mohamed Ouchrif}\\
\vspace{0.5cm}
{Dipartimento di Fisica dell'Universit\`a di Bologna\\
and INFN Sezione di Bologna, Viale Berti  Pichat 6/2, I-40127, Italy.\\
 and\\
Universit\'e Mohamed Premier, Facult\'e des Sciences,\\
        D\'epartement de Physique, L.P.T.P.
        B.P: 524 ~~60000~Oujda,~~Morocco.\\
        {\em E-mail: ouchrif@bo.infn.it} }
\end{center}

\vspace{2cm}
\begin{center}
{\bf Abstract}
\end{center}
\vspace{1cm}
We present a sudy of the interactions of Q-balls with matter, and  their
energy losses in the
earth, for a large range of velocities. These calculations are used to compute
the fractional geometrical acceptance of underground detectors.
Furthermore
 we computed the light yield in liquid
scintillators, the ionization in streamer tubes and the Restricted Energy Loss
 in nuclear track detectors.
\vspace{2cm}

\begin{center}
{\em Invited talk at COSMO99, ICTP-Trieste September 1999}
\end{center}

\newpage 
\small
{\small

\vspace{5.5cm}

\begin{center}

{\bf Energy losses of Q-balls in Matter, Earth and Detectors.}\\
\vspace{1.5cm}
{ Mohamed Ouchrif}\\
{Dipartimento di Fisica dell'Universit\`a di Bologna\\
and INFN Sezione di Bologna, Viale Berti  Pichat 6/2, I-40127, Italy.\\
 and\\
Universit\'e Mohamed Premier, Facult\'e des Sciences,\\
        D\'epartement de Physique, L.P.T.P.
        B.P: 524 ~~60000~Oujda,~~Morocco.\\
	{\em E-mail: ouchrif@bo.infn.it} }
        
\end{center}

We present a sudy of the interactions of Q-balls with matter, and  their
energy losses in the
earth, for a large range of velocities. These calculations are used to compute
the fractional geometrical acceptance of underground detectors. 
Furthermore 
 we computed the light yield in liquid
scintillators, the ionization in streamer tubes and the Restricted Energy Loss
 in nuclear track detectors.

\section{Interaction of Q-balls with matter}
\subsection{Interaction with matter of Q-balls type SECS}
SECS (Supersymmetric Electrically Charged Solitons) 
are Q-balls with a net positive electric charge which tends to be mainly 
in the outer layer. The 
charge of SECS originates from the unequal rates of absorption in the 
condensate$^{1}$. 
The positive  electric charge could be of one unit up to several tens. 
This 
positive electric charge may be
neutralized by a surrounding cloud of electrons.  

For small size Q-balls the positive charge 
interacts with matter (electrons and nuclei) via elastic 
or quasi elastic collisions$^{2}$. 
The cross section is similar to the Bohr cross section of hydrogenoid 
atoms$^{3-4}$
\begin{equation}
\sigma = \pi a_{0}^{2}\sim 10^{-16}cm^2 \label{eq:7}
\end{equation}
where $a_{0}$ is the Bohr radius.

The main energy losses$^{5}$ of SECS passing through 
matter with velocities in the 
range $10^{-4}<\beta<10^{-2}$ are  due to two contributions: 
$(i)$ the interaction of the SECS positive charge 
 with the nuclei (nuclear contribution),
and, $(ii)$ with the electrons of the traversed 
medium  (electronic contribution). 
The total energy loss is the sum of the two 
contributions$^{5-6}$.

{\bf Electronic energy losses of SECS:} The electronic$^{5-7}$ contribution 
to the energy 
loss of SECS may be computed
with the following formula
\begin{equation}
\frac{dE}{dx} = \frac{8 \pi a_{0} e^{2}  \beta }{\alpha} \frac{Z_{Q}^{7/6} N_{e}}{
(Z_{Q}^{2/3} + Z^{2/3})^{3/2}} 
\end{equation}
where $\alpha$ is the fine structure constant, $\beta = v/c$,  
$Z_{Q}$ is the positive charge of SECS, $Z$ is the atomic number 
of the medium and 
$N_e$ is the density of electrons in the medium. Electronic losses dominate 
for $\beta > 10^{-4}$ for the 
case of the dE/dx in the earth$^{5}$.

{\bf Nuclear energy losses of SECS:} The nuclear$^{5}$ contribution to 
the energy loss 
of SECS is due to the interaction of the SECS positive core 
with the nuclei of the medium and it is given by 
\begin{equation}
\frac{dE}{dx} = \frac{\pi a^{2} \gamma N E }{\epsilon} S_{n}(\epsilon) 
\label{eq:10}
\end{equation}
where
\begin{equation}
S_{n}(\epsilon) \simeq \frac{0.56 ~ln(1.2\epsilon)}{1.2\epsilon - 
(1.2\epsilon)^{-0.63}}~,~~~\epsilon = \frac{a M E}{Z_{Q}Z e^{2}M_{Q}}
\label{eq:11}
\end{equation}
and
\begin{equation}
a= \frac{0.885~ a_{0}}{(\sqrt{Z_{Q}} + \sqrt{Z})^{2/3}}~,~~~~~~~~~~~~
\gamma= \frac{4 M}{M_{Q}} \label{eq:12}
\end{equation}
$M_{Q}$ is the mass of the incident Q-ball; $M$ is the mass of the 
target nucleus; $Z_{Q}e$ and $Ze$ are their electric charges; we assume 
that $M_{Q} >> M$. Nuclear energy losses dominates for $\beta \leq 10^{-4}$.

Using the energy losses of SECS discussed above, 
we can computed,  
for a specific velocity 
$v = 250~km/s$, 
the angular  acceptance$^{6}$ of a detector located at the underground 
Gran Sasso Laboratory in Italy (MACRO experiment).

\subsection{Interaction with matter of Q-balls type SENS}
The Q-ball interior of SENS (Supersymmetric Electrically Neutral Solitons) 
is characterized by a 
large Vacuum Expectation Value  (VEV) of squarks,   
and may be of sleptons and Higgs fields. 
The $SU(3)_{c}$ symmetry is broken and deconfinement takes 
place inside the Q-ball. If 
a nucleon enters this region of deconfinement, it dissociates into three 
quarks, some of  
which may  then become absorbed in the supersymmetric condensate.
The reaction looks like$^{3}$
\begin{equation}
(Q) + Nucleon \rightarrow (Q+1) + pions \label{eq:13}
\end{equation}
or less probably, 
\begin{equation}
(Q) + Nucleon \rightarrow (Q+1) + kaons \label{eq:14}
\end{equation}

The nucleon enter the $\tilde{q}$ condensate and it gives rise to the process 
\begin{equation}
qq \rightarrow \tilde{q}\tilde{q}
\end{equation}

If it is assumed that the energy released in 
(\ref{eq:13}) and  (\ref{eq:14})   
is the same as in typical 
hadronic processes 
(about 1~GeV per nucleon), this energy is  carried by 2 or 3 pions (or 
kaons). 
The cross section 
 is determined by the Q-ball radius  
\begin{equation}
\sigma = \pi~R_{Q}^2 = \frac{16 \pi^2}{9}~M_{Q}^{-2}~Q^2 
\sim 6 \times 10^{-34}Q^{1/2} \left( \frac {1~TeV}{M_S} 
\right)^2~cm^2  \label{eq:15}
\end{equation}
The corresponding mean free path $\lambda$ is 
\begin{equation}
\lambda = \frac{1}{\sigma N}  \label{eq:16}
\end{equation}

The 
energy loss of SENS moving with velocities in the range 
$10^{-4}<\beta <10^{-2}$ is constant and is given by 
\begin{equation}
\frac{dE}{dx} \sim \frac{\zeta}{\lambda} = \sigma N \zeta
\end{equation}
where $N$ is the number of atoms per $cm^3$ in the traversed material;  
$\zeta = 1~(\frac{\rho}
{1g.cm^{-3}}) GeV$ is 
the energy released in one reaction.  
Large mass SENS lose a small fraction of their 
kinetic energy and are able to traverse the earth without attenuation for all 
masses of our interest.

\section{ Energy losses of Q-balls in the earth }
In general the energy losses in the earth are calculated using the 
density profile of the earth interior. One may observe 
 three layers: the nucleus, the mantle and the crust. 
For our purposes it is sufficient to use a simpler model, 
in which the density and composition of each layer is uniform. 

 In the earth interior model the nucleus is made of iron, with a density
of 11.5 $\mbox{g/cm}^{3}$ and a
conductivity of $1.6\times 10^{16}~{s}^{-1}$;
the mantle is made of Si, with a density of
4.3 $\mbox{g/cm}^{3}$. The radius of the nucleus is 0.54 earth
radii. The crust may be neglected as long as we consider Q-balls 
arriving at underground detectors 
from below. The rock above is assumed to have the same
 composition of the mantle.

The energy losses of SECS in the earth mantle and earth core can be easily 
computed for different $\beta$-ranges and for different positive electric 
charges of the Q-ball
core.

\section{Energy losses of SECS in detectors}
 
\subsection{Light Yield of SECS in Scintillators}
For SECS we distinguish two contributions to the light yield in scintillators: 
the primary light yield  and the secondary light yield.

{\bf \em The primary light yield}  is due to the direct excitation (and 
ionization that occurs only for $\beta > 10^{-3}$)
produced by the SECS in the medium. The energy loss in the MACRO 
liquid scintillator is computed from the energy loss of protons in hydrogen 
and carbon$^{5-8}$

{\bf *} For electric charge $q=1e$ the energy loss of SECS is 
calculated for two cases

{\bf i)}~For $10^{-5}  <   \beta < 5 \times 10^{-3}$ 
we have the following formula
\begin{equation}
\left(\frac{dE}{dx} \right)_{SECS}
= 1.3 \times 10^{5} \beta \;\; \left[~1- \exp \left( \frac{\beta}{
7 \times 10^{-4}} \right)^{2} \right]~~~\frac{MeV}{cm} \label{eq:19}
\end{equation}

{\bf ii)}~For $5 \times 10^{-3}  <   \beta < 10^{-2}$ we used the following 
formula  
\begin{equation}
SP = SP_{H} + SP_{C} = \left(\frac{dE}{dx} \right)_{SECS} \label{eq:20}
\end{equation}
where
\begin{equation}
SP_{H} = \frac{SL_{H} \times SH_{H}}{SL_{H} + SH_{H}} \label{eq:21}
\end{equation}

\begin{equation}
SP_{C} = \frac{SL_{C} \times SH_{C}}{SL_{C} + SH_{C}} \label{eq:22}
\end{equation}
and
\begin{equation}
SL = A_{1}~E^{0.45},~~~~~~~~~SH = A_{2}~Ln \left( 1 + 
\frac{A_{3}}{E} + A_{4} E \right) \label{eq:23}
\end{equation}
where ($A_{i=1,4}$) are constants obtained from experimental data, and $E$ is 
the energy of a proton with velocity $\beta$.

{\bf *} For SECS with electric charge $q = Z_{1}e$ the energy losses for $ 10^{-5} < 
\beta < 10^{-2}$
are given by$^{9-11}$
\begin{equation}
\left(\frac{dE}{dx} \right)_{SECS} = \frac{ 8 \pi e^{2} a_{0} \beta}
{\alpha} \frac{Z_{1}^{7/6} N_{e}}
{(Z_{1}^{2/3} + Z_{2}^{2/3})^{3/2}} \left[ ~1- \exp \left( - \frac{\beta}
{7 \times 10^{-4}} \right)^2 \right] \label{eq:24}
\end{equation}
where $Z_{2}$ is the atomic number of the target atom, $N_e$ the density 
of electrons  and $\alpha$ is the fine structure constant.

The primary light yield of SECS is given by 
\begin{equation}
\left(\frac{dL}{dx} \right)_{SECS} 
= A~ \left[ \frac{1}{1 + AB~\frac{dE}{dx}} \right]~\frac{dE}{dx} \label{eq:25}
\end{equation}
where $dE/dx$ is the total energy loss of SECS;   
$A$ is a constant of conversion of the energy losses 
in photons (light yield) and $B$ is the 
parameter describing the saturation of the light yield; both  
parameters depend only on the velocity of SECS. 

{\bf \em The secondary light yield} arises from recoiling particles: 
we consider the elastic or quasi-elastic
recoil of hydrogen and carbon nuclei. The light yield $L_{p}$ from  
a hydrogen or carbon nucleus of
given initial energy $E$ is computed as

\begin{equation}
L_{p}(E)=\int^{E}_{0}\frac{dL}{dx}(\epsilon)S^{-1}_{tot} \, d\epsilon
\label{eq:26}
\end{equation}
where $S_{tot}$ is the sum of electronic and nuclear 
energy losses.
The secondary light
yield is then
\begin{equation}
\left(\frac{dL}{dx}\right)_{\mbox{secondary}}=N\int^{T_{m}}_{0}L_{p}(T)
\frac{d\sigma}{dT} \, dT \label{eq:27}
\end{equation}
where $N$ is the number density of atoms in the medium  
$T_{m}$ is the maximum energy transferred and
 {\Large $\frac{d\sigma}{dT}$} is the
differential scattering cross section.

\subsection{Energy losses of SECS in streamer tubes}
The composition of the gas in the  MACRO limited streamer tubes is 73\% 
helium and   27\% n-pentane, in volume.  
The pressure is about one atmosphere and the resulting density
is low (in comparison with the density of the
other detectors): $ \rho_{gas}=0.856~\mbox{mg/cm}^{3}$.

The ionization energy losses of SECS with $10^{-3} < \beta < 10^{-2}$ 
in the MACRO streamer tubes are computed with the same general 
procedure used 
for scintillators, using the density and the chemical composition of 
streamer tubes. 

The threshold for ionizing n-pentane occurs for  $\beta \geq 2 \times 10^{-3}$.

\subsection{Restricted Energy Losses of SECS in the 
Nuclear Track Detectors}
The relevant parameter for nuclear track detectors is the
Restricted Energy Loss (REL), that is, the energy deposited within $\sim$~
100~\AA~~ from the track.

The chemical composition of CR39 nuclear track detector is
$(\mbox{C}_{12}\mbox{H}_{18}\mbox{O}_{7})_n$, and the
density is 1.31~$\mbox{g/cm}^3$.
For the computation of the REL, only energy transfers to atoms 
larger than $12$~eV are
considered, because it is estimated that $12$~eV are necessary to break the
molecular bonds.

At {\em low velocities} ($3 \times 10^{-5}<\beta<10^{-2}$) there are
two contributions to REL: the ionization and the atomic recoil 
contributions.

{\em The ionization contribution,} which become important only for $\beta > 
2 \times 10^{-3}$,   
was computed
with Ziegler's fit to the experimental data. 

{\em The atomic recoil contribution,}  important for low $\beta$ values,  
and was calculated  
using the  interaction potential between an atom and a SECS which is  

\begin{equation}
V(r) = \frac{Z_{1} Z_{2} e^{2}}{r} \phi (r) \label{eq:28}
\end{equation}
where $r$ is the distance between the core of SECS and the target atom, 
$Z_{1}e$ is 
the electric charge of the SECS core,  
$Z_{2}$ is the atomic number of the target atom. 

The Restricted 
Energy Losses are finally obtained by
integrating the transferred energies as
\begin{equation}
-\frac{dE}{dx}  = N \int \sigma(K) \, dK \label{eq:34}
\end{equation}
where $N$ is the number density of atoms in the medium, $\sigma(K)$ is the
differential cross section as function of the transferred kinetic energy K.

\section{Conclusions}

We computed for a large range of velocity the energy losses of Q-balls 
of type SENS and SECS in matter.               
Using these energy losses and a rough model of the earth's  composition 
and density profiles, we have computed the energy losses in the earth interior. 

We also calculated  the energy deposited in 
scintillators, streamer tubes and nuclear track detectors by SECS, in 
forms useful for their detection. In particular we computed the light yield 
in scintillators, the ionization in streamer tubes and the 
REL in nuclear track detectors.

}


\newpage
\begin{thebibliography}{99}
\bibitem{TDLEE92} T.D. Lee and Y. Pang, Phys. Rept. {221} (1992);\\
S. Coleman, Nucl. Phys. {B262} (1985) 293.\\
\vspace{-0.5cm}
\bibitem{Kusenko97} A. Kusenko, Phys. Lett. { B405} (1997) 108;\\ 
Phys. Lett. {B404} (1997) 285;\\
Phys. Lett. {B406} (1997) 26.\\
\vspace{-0.5cm}
\bibitem{Kusenko98A} A. Kusenko and M. Shaposhnikov, Phys. Lett. 
{\bf B 417} (1998) 99;\\
A. Kusenko et al. Phys. Rev. Lett. {80} (1998) 3185.\\
\vspace{-0.5cm}
\bibitem{Ouchrif98} M. Ouchrif, PhD. thesis, University of Oujda (1999) .\\
\vspace{-0.5cm}
\bibitem{Kusenko98B} D. Bakari et al., {`` \em Energy Losses of Q-balls."},  
[hep-ex/{0003003}], submitted to Astroparticle Physics (2000).\\
\vspace{-0.5cm}
\bibitem{OUCHRIF} M. Ouchrif, {``\em Q-balls in Underground Experiments"}, 
[hep-ex/{0002013}].\\
\vspace{-0.5cm}
\bibitem{Kolda} T. Gherghetta, C. Kolda and S.P. Martin, Nucl. Phys. 
{\bf B468} (1996) 37.\\
\vspace{-0.5cm}
\bibitem{Z77} H. H. Andersen and J.F. Ziegler, 
{``\em Hydrogen stopping power and ranges in all elements"}, Pergamon Press
(1977). \\
\vspace{-0.5cm}
\bibitem{F87} D. J. Ficenec et al., Phys. Rev. {D36} (1987) 311. \\
\vspace{-0.5cm}
\bibitem{L61} J. Lindhard and M. Scharff, Phys. Rev. {124}
 (1961) 28.\\
\vspace{-0.5cm}
\bibitem{D97} J. Derkaoui al., Astropart. Phys. {10} (1999) 339;\\
Astropart. Phys. {9} (1998) 173.\\
\vspace{-0.5 cm}
\bibitem{W77} W. D. Wilson, L. G. Haggmark and J. P. Biersack,
Phys. Rev. {B15}  (1977) 2458. \\
\vspace{-0.5cm}
\bibitem{L77} J. Lindhard et al.,
K. Dan. Vidensk. Selsk. Mat.-Fys. Med. {33}  (1963) No. 14.\\
\vspace{-0.5cm}
\bibitem{R83} T. W. Ruijgrok, J. A. Tjon and T. T. Wu, Phys. Lett. { 129B}
 (1983) 209; \\
S. Nakamura, PhD. Thesis,  UT-ICEPP-88-04, University of Tokyo (1988).\\ 


\end{thebibliography}
\end{document}